\documentclass[10pt,twocolumn,letterpaper]{article}

\usepackage{cvpr}
\usepackage{times}
\usepackage{epsfig}
\usepackage{graphicx}
\usepackage{amsmath}
\usepackage{amssymb}

\usepackage{wrapfig}
\usepackage{subfigure}

\usepackage{algorithm}
\usepackage{algorithmic}

\usepackage[pagebackref=true,breaklinks=true,letterpaper=true,colorlinks,bookmarks=false]{hyperref}

\usepackage[breaklinks=true,bookmarks=false]{hyperref}

\cvprfinalcopy 

\def\cvprPaperID{****} 

\begin{document}

\title{ Globally optimal point set registration by joint symmetry plane fitting}

\author{Lan Hu, Haomin Shi, and Laurent Kneip\\Mobile Perception Laboratory, ShanghaiTech University\\  \{hulan, shihm, lkneip\}@shanghaitech.edu.cn
}
\maketitle
\begin{abstract}
The present work proposes a solution to the challenging problem of registering two partial point sets of the same object with very limited overlap. We leverage the fact that most objects found in man-made environments contain a plane of symmetry. By reflecting the points of each set with respect to the plane of symmetry, we can largely increase the overlap between the sets and therefore boost the registration process. However, prior knowledge about the plane of symmetry is generally unavailable or at least very hard to find, especially with limited partial views, and finding this plane could strongly benefit from a prior alignment of the partial point sets. We solve this chicken-and-egg problem by jointly optimizing the relative pose and symmetry plane parameters, and notably do so under global optimality by employing the branch-and-bound (BnB) paradigm. Our results demonstrate a great improvement over the current state-of-the-art in globally optimal point set registration for common objects. We furthermore show an interesting application of our method to dense 3D reconstruction of scenes with repetitive objects.
\end{abstract}

\section{Introduction}
The alignment of two point sets is a fundamental geometric problem that occurs in many computer vision and robotics applications. In computer vision, the technique is used to stitch together partial 3D reconstructions in order to form a more complete model of an object or environment \cite{pulli1999multiview}. In robotics, point set registration is an essential ingredient to simultaneous localization and mapping with affordable consumer depth cameras~\cite{newcombe11} or powerful 3D laser range scanners \cite{chen20133d}. The general approach for aligning two point sets does not require initial correspondences. It is given by the Iterative Closest Point (ICP) algorithm~\cite{zhang1994iterative}, a local search strategy that alternates between geometric correspondence establishment (i.e. by simple nearest neighbour search) and Procrustes alignment. The iterative procedure depends on a sufficiently accurate initial guess about the relative transformation (e.g. an identity transformation in the case of incremental ego-motion estimation).

The present work is motivated by a common problem that occurs when performing a dense reconstruction of an environment which contains multiple instances of the same object. An example of the latter is given by a room in which the same type of chair occurs more than once. Let us assume that the front-end of our reconstruction framework encompasses semantic recognition capabilities which are used to segment out partial point sets of objects of the same class and type~\cite{maccormac18}. There is a general interest in aligning those partial object point sets towards exploiting their mutual information and completing or even improving the reconstruction of each instance. The difficulty of this partial point set registration problem arises from two factors:
\begin{itemize}
\item The relative pose between the different objects is arbitrary and unknown.
\item Since the objects are observed under an arbitrary pose and with potential occlusions, the measured partial point sets potentially have very little overlap.
\end{itemize}

Our contribution focuses on the registration of only two partial point sets, which appears as a worthwhile starting point given that the registration of more than two point sets may be broken down into many pair-wise alignments. The plain ICP algorithm is only a local search strategy that depends on a sufficiently accurate initial guess, which is why it may not serve as a valid solution to our problem. A potential remedy is given by the globally optimal ICP algorithm presented by Yang et al. \cite{yang2015go}. However, the algorithm still depends on sufficient overlap in the partial point sets, which is not necessarily a given (50\% is reported as a requirement for high success rate).

The core idea of the present work consists of exploiting the fact that the majority of commonly observed objects contain a plane of symmetry. By reflecting the points of each partial point set with respect to the plane of symmetry, we may effectively increase the overlap between the two sets and vastly improve the success rate of the registration process. However, given that each point set only observes part of the object, the identification of the plane of symmetry in each individual point set appears to be an equally difficult and ill-posed problem than the partial point set registration problem itself. It is only after the aligning transformation is found that symmetry plane detection would become a less challenging problem. In conclusion, the solution of each problem strongly depends on a prior solution to the other. We therefore present the following contributions:
\begin{itemize}
\item We solve this chicken-and-egg problem by a joint solution of the aligning transformation as well as the symmetry plane parameters. To the best of our knowledge, our work is the first to address those two problems jointly, thus leading to a vast improvement over the existing state-of-the-art in globally optimal point set registration, especially in situations in which two point sets contain very limited overlap.
\item We immediately provide a globally optimal solution to this problem by employing the branch-and-bound optimisation paradigm. Our work implicitly provides the first solution to globally optimal symmetry plane estimation in a single point set, or---more generally---symmetry plane detection across two point sets.
\item We show a meaningful application of our algorithm in a dense 3D reconstruction scenario in which multiple instances of the same object occur. 
\end{itemize}

\section{Related Work}

Despite strong mutual dependency, our method is the first to perform joint point set alignment and symmetry plane estimation. Our literature review therefore cites prior art on those two topics individually.


\subsection{Point-set Registration}


The Iterative Closest Point (ICP) \cite{chen1992object, besl1992method, zhang1994iterative} algorithm is a popular method for aligning two point sets. It does not depend on a prior derivation of point-to-point correspondences, and simply aligns the two sets by iteratively alternating between the two steps of finding nearest neighbours (e.g. by minimising point-to-point distances), and computing the alignment (e.g. using Arun's method~\cite{arun87}). To improve robustness of the algorithm against occlusions and reduced overlap, the method has been extended by outlier rejection \cite{zhang1994iterative,granger2002multi} or data trimming \cite{chetverikov2005robust} techniques. However, the classical ICP algorithm remains a local search algorithm for which the convergence depends on initial guess and sufficient overlap between both point sets.

An entire family of alternative approaches relies on the idea of expressing both point sets by a Gaussian Mixture Model (GMM) and aligning the latter using Gaussian Mixture Alignment (GMA). Notable GMA-based techniques for rigid and non-rigid registration are given by the robust point matching algorithm by Chui and Rangarajan \cite{chui2003new}, the coherent point drift strategy by Myronenko and Song \cite{myronenko2010point}, kernel correlation by Tsin and Kanade \cite{tsin2004correlation}, and the GMMReg algorithm by Jian and Vemuri \cite{jian2010robust}. While GMA is advertised by improved robustness against poor initialisations, noise, and outliers, another key advantage with respect to point-based methods is given by a closed-form expression to evaluate the quality of the alignment (i.e. the method does not depend on an alternating search for nearest neighbours). However, the listed GMA algorithms remain local search algorithms, which makes them inapplicable in scenarios in which no prior about the relative transformation is given upfront.

In contrast, globally optimal algorithms avoid local minima by searching the entire space of relative transformations, often using the branch-and-bound paradigm \cite{li20073d, olsson2008branch, parra2014fast}. Yang et al. \cite{yang2013go,yang2015go} propose the Go-ICP algorithm, which applies  branch-and-bound to the ICP problem to find the globally optimal minimum of the sum of L2-distances between nearest neighbours from two aligned point sets. The method is accelerated by using local ICP in the loop. However, missing robustness of the cost-function causes the method to remain sensitive with respect to occlusions and partial overlaps. Campbell et al. \cite{campbell2016gogma} finally devise GOGMA, a branch-and-bound variant in which the objective of minimising point-to-point errors is again replaced by the convolution of GMMs. 

Inspired by those recent advances, we also employ the nested branch-and-bound strategy integrated with local ICP to find a globally optimal alignment. However, in contrast to all prior art, we are the first to jointly fit a plane of symmetry, which leads to a large improvement in partial scan alignment for common symmetric objects.

\subsection{Symmetry Plane Estimation}

For a comprehensive review of symmetry detection, we kindly refer the reader to Liu et al.'s review \cite{liu2010computational}. Here we only focus on the problem of symmetry plane fitting with missing data. The most straightforward solution is given by employing the RANdom SAmple Consensus (RANSAC) algorithm proposed by Fischler and Bolles \cite{fischler1981random}, a well-known algorithm for robust model fitting for outlier affected data. In the context of shape matching, the basic idea is to extract sparse characteristic points and match them between both sets. We then choose a random subset of correspondences and derive a hypothesis for the global transformation induced by these samples.
The alignment quality is finally evaluated by the matching error between the two shapes. The method can be easily applied for detecting a plane of symmetry in a single point set by hypothesising the latter to be orthogonal to the axis connecting a correspondence. For example, Cohen et al. \cite{cohen2012discovering} detect symmetries in sparse point clouds by using appearance-based 3D-3D point correspondences in a RANSAC scheme. The detected symmetries are subsequently explored to eliminate noise from the point-clouds. Xu et al. \cite{xu2009partial} present a voting algorithm to detect the intrinsic reflectional symmetry axis. Using the axis as a hint, a completion algorithm for missing geometry is again shown. Jiang et al. \cite{jiang2013skeleton} on the other hand propose an algorithm to find intrinsic symmetries in point clouds by using a curve skeleton. A set of filters then produces a candidate set of symmetric correspondences which are finally verified via spectral analysis. Although these works show results on partial data, the amount of missing data is typically small. Inspired by the work of \cite{cicconet2017finding} which detects symmetry by registration, we propose to detect the symmetry plane alongside partial point set registration, thus leading to improved performance in situations with limited overlap.

\section{Preliminaries}

We start by introducing the notation used throughout the paper and review the basic formulation of the ICP problem as well as planar reflections.

\subsection{Notations and Assumptions}

Let us denote the two partial object point sets by $\mathcal{X} = \{\mathbf{x}_i\}_{i=1}^{M}$ and $\mathcal{Y} = \{\mathbf{y}_i\}_{i=1}^{N}$ (sometimes called the \emph{model} and \emph{data} point sets, respectively). The goal pursued in this paper is the identification of a Euclidean transformation given by the rotation $\mathbf{R}$ and the translation $\mathbf{t}$ that transforms the points of $\mathcal{Y}$ such that they \textit{align} with the points of $\mathcal{X}$. If $\mathcal{X}$ and $\mathcal{Y}$ contain points in the 2D plane, $\mathbf{R}$ and $\mathbf{t}$ form an element of the group $SE(2)$. If $\mathcal{X}$ and $\mathcal{Y}$ contain 3D points, $\mathbf{R}$ and $\mathbf{t}$ will be an element of $SE(3)$. Note that \textit{alignment} denotes a more general idea rather than just the minimisation of the sum of distances between each point of $\mathcal{Y}$ and its closest point within $\mathcal{X}$. The point sets have different cardinality and potentially observe very different parts of the object with only very little overlap. This motivates our approach that takes object symmetries into account.

\subsection{Registration of Two Point Sets}

The standard solution to the point set registration problem is given by the ICP algorithm~\cite{zhang1994iterative}, which minimizes the alignment error given by
\begin{equation}
E(\mathbf{R},\mathbf{t}) = \sum_{i=1}^{N}e^r_i(\mathbf{R},\mathbf{t}|\mathbf{y}_i)^2 = \sum_{i=1}^{N} \|\mathbf{R}  \mathbf{y}_i+\mathbf{t}-\mathbf{x}_{j}\|^2,
\label{eq:procrustes}
\end{equation}
where $e^r_i(\mathbf{R},\mathbf{t}|\mathbf{y}_i)$ is the per-point residual error for $\mathbf{y}_i$, and $\mathbf{x}_{j}$ is the closest point to $\mathbf{y}_i$ within $\mathcal{X}$, i.e.
\begin{equation}
\mathbf{x}_{j} = \underset{\mathbf{x}\in\mathcal{X}}{\operatorname{argmin}}{\|\mathbf{R}\mathbf{y}_i+\mathbf{t}-\mathbf{x}\|}.
\label{eq:nearestNeighbours}
\end{equation}
Given an initial transformation $\mathbf{R}$ and $\mathbf{t}$, the ICP algorithm iteratively solves the above minimization problem by alternating between updating the aligning transformation with fixed $\mathbf{x}_{j}$ (i.e. using \eqref{eq:procrustes}), and updating the closest-point matches $\mathbf{x}_{j}$ themselves using \eqref{eq:nearestNeighbours}. It is intuitively clear that the ICP algorithm only convergence to a local minimum.

\subsection{Modelling and Identifying Symmetry}

Symmetry is modelled by a reflection by the symmetry plane. Let us define the symmetry plane by the normal $\mathbf{n}$ and depth-of-plane $d$ such that any point $\mathbf{x}$ on the plane satisfies the relation $\mathbf{x}^T\mathbf{n}+d = 0$. Let $\mathbf{x}$ now be a (2D or 3D) point from $\mathcal{X}$. The reflection plane reflects $\mathbf{x}$ to a single reflected point $\hat{\mathbf{x}}$ given by
\begin{equation}
\hat{\mathbf{x}} = \mathbf{x} - 2\mathbf{n}( \mathbf{x}^T\mathbf{n} + d).
\label{eq:reflection}
\end{equation}
The term in parentheses is the signed distance between $\mathbf{x}$ and the reflection plane. The subtraction of $2\mathbf{n}$ times this distance reflects the point to the other side of the plane. The problem of symmetry identification may now be formulated as a minimisation of the symmetry distance defined by
\begin{equation}
E(\mathbf{n},d) = \sum_{i=1}^{M}e^s_i(\mathbf{n},d|\mathbf{x}_i)^2 = \sum_{i=1}^{M} \|\hat{\mathbf{x}}_i-\mathbf{x}_j\|^2,
\label{eq:symmetricObjective1}
\end{equation}
where $e^s_i(\mathbf{n},d|\mathbf{x}_i)$ is the per-point residual error for $\mathbf{x}_i$, and $\mathbf{x}_j$ is the nearest point to $\hat{\mathbf{x}}_i$ in $\mathcal{X}$, i.e.
\begin{equation}
\mathbf{x}_{j} = \underset{\mathbf{x}\in\mathcal{X}}{\operatorname{argmin}}{\|\hat{\mathbf{x}}_i-\mathbf{x}\|}.
\label{eq:nearestNeighbours2}
\end{equation}
It is intuitively clear that the symmetry plane fitting problem may also be solved via ICP, the only difference being the parameters over which the problem is solved (i.e. $\mathbf{n}$ and $d$).

\subsection{Transformed Symmetry Plane Parameters}

Let us still assume that  $\mathbf{n}$ and $d$ are the symmetry plane parameters of a point set $\mathcal{X}$, and $\mathbf{R}$ and $\mathbf{t}$ are the parameters that align a point set $\mathcal{Y}$ with $\mathcal{X}$. Each transformed point $\mathbf{R}\mathbf{y}+\mathbf{t}$ and its transformed, symmetric equivalent $\mathbf{R}\hat{\mathbf{y}}+\mathbf{t}$ must still fulfill the original reflection equation \eqref{eq:reflection}:
\begin{equation}
\mathbf{R}\hat{\mathbf{y}}+\mathbf{t} = \mathbf{R}\mathbf{y}+\mathbf{t}  - 2\mathbf{n}( (\mathbf{R}\mathbf{y}+\mathbf{t})^T\mathbf{n} + d).
\end{equation}
Cancelling $\mathbf{t}$ and multiplying by $\mathbf{R}^T$ on either side, we easily obtain
\begin{equation}
\hat{\mathbf{y}} = \mathbf{y}  - 2\mathbf{R}^T\mathbf{n}( \mathbf{y}^T \mathbf{R}^T\mathbf{n} +\mathbf{t}^T \mathbf{n}+d).
\label{eq:reflection_transformation}
\end{equation}
By comparing to (\ref{eq:reflection}), it is obvious that  $\hat{\mathbf{n}} = \mathbf{R}^T\mathbf{n}$ and $\hat{d} = \mathbf{t}^T\mathbf{n}+d$ must represent the symmetry plane parameters for the original, untransformed set $\mathcal{Y}$.
 
\section{Alignment and Symmetry as a Joint Optimization Problem}

We now introduce our novel optimization objective which jointly optimizes an aligning point set transformation as well as the plane of symmetry. The objective is then solved in a branch-and-bound optimization paradigm, for which we introduce both the domain parameterization as well as the derivation of upper and lower bounds.

\subsection{Objective Function}

We still assume that our two partial object point sets are given by $\mathcal{X} = \{\mathbf{x}_i\}_{i=1}^{M}$ and $\mathcal{Y} = \{\mathbf{y}_i\}_{i=1}^{N}$, and that the symmetry plane is represented by the normal $\mathbf{n}$ and depth $d$. We define $\mathcal{X}^s = \{\mathbf{x}^s_i | \mathbf{x}_i^s = \mathbf{x}_i-2 \mathbf{n}(\mathbf{x}_i^T\mathbf{n}+d), i=1,\cdots,M \}$ to be the corresponding reflected point set of $\mathcal{X}$. We furthermore define $\mathcal{X}^r = \{\mathbf{x}^r_i | \mathbf{x}_i^r = \mathbf{R}^T(\mathbf{x}_i-\mathbf{t}), i=1,\cdots,M \}$ and $\mathcal{Y}^r = \{\mathbf{y}^r_i | \mathbf{y}_i^r = \mathbf{R}\mathbf{y}_i+\mathbf{t}, i=1,\cdots,N \}$ to be the aligned sets in either direction. The symmetry fitting objective of $\mathcal{X}$ employs
\begin{equation}
e^s_i(\mathbf{n},d| \mathbf{x}_i) =  \|\mathbf{x}_i - 2\mathbf{n}( \mathbf{x}_i^T\mathbf{n} + d)-\mathbf{x}_j \|,
\label{eq:reflection-model-error-term}
\end{equation}
where the difference to \eqref{eq:symmetricObjective1} is given by the fact that $\mathbf{x}_j$ is now the nearest neighbour from the set $\mathcal{X} \bigcup \mathcal{Y}^r$. Similarly, using equation (\ref{eq:reflection_transformation}), the symmetry objective function for $\mathcal{Y}$ employs
\begin{equation}
e^s_i(\hat{\mathbf{n}}, \hat{d}|\mathbf{y}_i) =  \|\mathbf{y}_i - 2\hat{\mathbf{n}}( \mathbf{y}_i^T\hat{\mathbf{n}} + \hat{d})-\mathbf{y}_j \|,
\end{equation}
where $\mathbf{y}_j$ is now chosen as the nearest neighbour from the set $\mathcal{X}^r \bigcup \mathcal{Y}$. The final registration error itself employs 
\begin{equation}
e^r_i(\mathbf{R},\mathbf{t}| \mathbf{y}_i) = w_i\|\mathbf{R}  \mathbf{y}_i+\mathbf{t}-\mathbf{x}_j\|,
\label{eq:registration-term}
\end{equation}
where $\mathbf{x}_j$ is chosen as the nearest neighbour from the set $\mathcal{X} \bigcup \mathcal{X}^s$ and weight $w_i$ is used to take the set $\mathcal{X}$ as aligned points first. The overall objective function becomes
\begin{eqnarray}
  E(\mathbf{R,t,n},d) & = & \sum_{i=1}^{M}e^s_i(\mathbf{n},d| \mathbf{x}_i)^2 \\ & + & \sum_{i=1}^{N} \left\{ e^r_i(\mathbf{R},\mathbf{t}| \mathbf{y}_i)^2+e^s_i(\hat{\mathbf{n}}, \hat{d}|\mathbf{y}_i)^2\right\}.\nonumber
  \label{eq:objective-function}
\end{eqnarray}
Direct optimization over $\mathbf{R,t,n}$, and $d$ using traditional ICP would easily get trapped in the nearest local minimum. We therefore propose to minimize the energy objective using the globally optimal branch-and-bound paradigm, an exhaustive search strategy that branches over the entire parameter space. In order to speed up the execution, the method derives upper and lower bounds for the minimal energy on each branch (i.e. sub-volume of the optimization space), and discards branches for which the lower bound remains higher than the upper bound in another branch. In the remainder of this section, we will discuss the two important questions of (i) how to parametrize and branch the parameter space, and (ii) how to find concrete values for the upper and lower bounds.
\subsection{Domain Parameterization}
\label{sec:domains}
\textbf{2D problem}: For a 2D point set, the domain of all feasible alignment parameters is represented by an angle $r$ to parameterize the planar rotation $\mathbf{R}_r = \begin{bmatrix} \cos(r) &-\sin(r) \\ \sin(r)& \cos(r) \end{bmatrix}$,
and a 2D translation vector $\mathbf{t}$.
The space of all 2D rotations can therefore be compactly represented by the interval $[-\pi,+\pi]$.
For the translation, we assume that the optimal translation must lie within a cube defined by the interval $[-\epsilon,+\epsilon ]^2$.
The symmetry plane in the 2D case becomes a line which is parameterized by an angle $\alpha$ defining the normal vector $\mathbf{n} = [\cos(\alpha) \text{ }\sin(\alpha)]^T$, and a scale $d$. Given the dual representation of line normal vectors, $\alpha$ is constrained to the interval $[-\frac{\pi}{2},+\frac{\pi}{2}]$, and $d$ lies in the interval $[-\varepsilon,+\varepsilon]$.

\textbf{3D problem}: The disadvantage of BnB is that its complexity grows exponentially in the dimensionality of the problem. We therefore take prior information about the 3D point sets into account that helps to decrease the dimensionality. We make the assumption that most objects are standing upright to the ground plane. The rotation between the different partial point sets is therefore still constrained to be a 1D rotation about the vertical axis, and the normal vector $\mathbf{n}$ remains a 1D variable the horizontal plane. The translation however becomes a 3D vector over the interval $[-\epsilon,+\epsilon ]^3$.

In conclusion, the 2D alignment problem has 5 degrees of freedom, whereas the 3D problem turns into a $6$-dimensional estimation problem.
\subsection{Derivation of the Upper and Lower Bounds}
The basic idea of BnB  is to partition feasible set into convex sets which means it crucially depends on upper and lower bounds for the objective $L$2-energy on a given interval. Given that our objective \eqref{eq:objective-function} consists of a sum of many squared and positive sub-energies, lower and upper bounds on the overall objective energy can be derived by calculating individual upper and lower bounds on the alignment and symmetry residuals. Using the above introduced parametrizations, the upper bound $\overline{E}$ and the lower bound $\underline{E}$ of the optimal, joint $L_2$ registration and symmetry cost $E^*$ on a given interval of variables are therefore given as
\small
\begin{eqnarray}
	\overline{E} & \doteq &  \sum_{i=1}^{M} \overline{e^s_i}(\alpha,d| \mathbf{x}_i)^2 \nonumber \\
	& + & \sum_{i=1}^{N} \{\overline{e^r_i}(r,\mathbf{t}| \mathbf{y}_i)^2+\overline{e^s_i}(r,\mathbf{t},\alpha,d| \mathbf{y}_i)^2 \} \\
	\underline{E} & \doteq &  \sum_{i=1}^{M} \underline{e^s_i}(\alpha,d| \mathbf{x}_i)^2 \nonumber \\
	& + & \sum_{i=1}^{N} \{\underline{e^r_i}(r,\mathbf{t}| \mathbf{y}_i)^2+\underline{e^s_i}(r,\mathbf{t},\alpha,d| \mathbf{y}_i)^2 \}
\end{eqnarray}
\normalsize
Upper bounds on an interval are easily given by the energy of an arbitrary point within the interval. Given an interval centered at $\{r_0,\mathbf{t}_0,\alpha_0,d_0\}$, upper bounds are therefore easily defined as
\small
\begin{eqnarray}
  \overline{e^s_i}(\alpha,d| \mathbf{x}_i) & = & e^s_i(\alpha_0,d_0| \mathbf{x}_i) \nonumber \\
  \overline{e^r_i}(r,\mathbf{t}| \mathbf{y}_i) & = & e^r_i(r_0,\mathbf{t}_0| \mathbf{y}_i) \nonumber \\
  \overline{e^s_i}(r,\mathbf{t},\alpha,d| \mathbf{y}_i) & = & e^s_i(r_0,\mathbf{t}_0,\alpha_0,d_0 | \mathbf{y}_i).
\label{eq:upperBounds}
\end{eqnarray}
\normalsize
The remainder of this section discusses the derivation of the lower bounds.

\textbf{Lower bound for the alignment error} $\underline{e^r_i}(r,\mathbf{t}| \mathbf{y}_i)$: For a rotation interval of half-length $\sigma_r$ with centre $r_0$, we have
\small
\begin{equation}
	\| \mathbf{R}_{r}\mathbf{y} -  \mathbf{R}_{r_0}\mathbf{y}\| \leq 2 \text{sin}(\text{min}(\sigma_r/2, \pi/2)) \|\mathbf{y}\|  \doteq \gamma_r \|\mathbf{y}\|.
	\label{eq:rotation-uncertainty}
\end{equation}
\normalsize
$\gamma_r$ is also called the rotation uncertainty radius.

\emph{Proof}: Using Lemmas 3.1 and 3.2 of \cite{hartley2009global}, we have:
\small
\begin{equation}
	\begin{split}
	 &\| \mathbf{R}_{r}\mathbf{y}-  \mathbf{R}_{r_0}\mathbf{y}\|\\
	 & = 2\text{sin}(\angle(\mathbf{R}_{r}\mathbf{y},  \mathbf{R}_{r_0}\mathbf{y} )/2)  \|\mathbf{y}\| \\
	 & \leq 2\text{sin}(\text{min}(\angle(\mathbf{R}_{r},  \mathbf{R}_{r_0}), \pi )/2)  \|\mathbf{y}\| \\
	 & \leq 2\text{sin}(\text{min}( |r-r_0|, \pi )/2)  \|\mathbf{y}\| \\
	 & \leq 2\text{sin}(\text{min}( \sigma_r/2, \pi/2 ))  \|\mathbf{y}\|.
	\end{split}
\end{equation}
\normalsize

We can similarly derive a translation uncertainty radius $\gamma_{{t}}$. For a translation volume with half side-length $\sigma_t$ centered at $\mathbf{t}_0$, we have
\small
\begin{equation}
	\|(\mathbf{x}+\mathbf{t}) -(\mathbf{x}+\mathbf{t}_0)\|= \|\mathbf{t}-\mathbf{t}_0\| \leq \sqrt{3}\sigma_t \doteq \gamma_t
\end{equation}
\normalsize
Note that in the 2D case, we have $\gamma_t \doteq \sqrt{2}\sigma_t$. The lower bound of the registration term in equation (\ref{eq:registration-term}) becomes 
\small
\begin{equation}{\label{eq:registration-lower}}
e^r_i(r,\mathbf{t}| \mathbf{y_i}) \geq  e^r_i(r_0,\mathbf{t}_0| \mathbf{y_i})-w_i(\gamma_r\|\mathbf{y}_i\|+\gamma_t) \doteq \underline{e^r_i}(r,\mathbf{t}| \mathbf{y}_i)
\end{equation}
\normalsize
For more details, we kindly refer the reader to \cite{hartley2009global,yang2015go}.

\textbf{Lower bound of symmetry term} $\underline{e^s_i}(\alpha,d| \mathbf{x}_i)$:
Assuming that the normal is defined by an $\alpha$-interval of half-length $\sigma_{\alpha}$ and with center $\alpha_0$, we have
\small
\begin{eqnarray}
	& &\|\mathbf{x}^T(\mathbf{n} -\mathbf{n}_0)\| \leq   \|\mathbf{n} -\mathbf{n}_0\|\|\mathbf{x}\| \nonumber\\
	&=&\sqrt{ \left(1-cos(\alpha-\alpha_0)\right)^2 + sin(\alpha-\alpha_0)^2}\|\mathbf{x}\| \nonumber\\ & \leq & \sqrt{ 2(1-\cos(\sigma_\alpha)) }\|\mathbf{x}\|
	\doteq \gamma_\alpha \|\mathbf{x}\|.
\end{eqnarray}
\normalsize
For the depth $d\in[d_0-\sigma_d,d_0+\sigma_d]$, we simply have
\small
\begin{equation}
	|d-d_0| \leq \sigma_d \doteq \gamma_d.
\end{equation}
\normalsize
Now let $\mathbf{x}_j \in \mathcal{X}\cup \mathcal{Y}^r$ be the closest point to $(\mathbf{x}_i - 2\mathbf{n}( \mathbf{x_i^T}\mathbf{n} + d))$ , and let $\mathbf{x}_{j_0} \in \mathcal{X}\cup \mathcal{Y}^r$ be the closest point to $(\mathbf{x}_i - 2\mathbf{n}_0( \mathbf{x}_i^T\mathbf{n}_0 + d_0))$. The lower bound is derived as follows:
\small
\begin{equation}
	\begin{split}
	   & e^s_i(\alpha,d| \mathbf{x}_i)
       = \|\mathbf{x}_i - 2\mathbf{n}( \mathbf{x}_i^T\mathbf{n} + d)-\mathbf{x}_j \| \\
	   & =\|\mathbf{x}_i - 2\mathbf{n}_0( \mathbf{x}_i^T\mathbf{n}_0 + d_0)-\mathbf{x}_j  \\
	   &\hspace{2cm}-2\left(\mathbf{n}( \mathbf{x}_i^T\mathbf{n}+d)-\mathbf{n}_0( \mathbf{x}_i^T\mathbf{n}_0 + d_0) \right) \| \\
	   &\geq \|\mathbf{x}_i - 2\mathbf{n}_0( \mathbf{x}_i^T \mathbf{n}_0 + d_0)-\mathbf{x}_{j_0}\| \\
	   &\hspace{2cm}-2\|\mathbf{n}( \mathbf{x}_i^T\mathbf{n}+d)-\mathbf{n}_0( \mathbf{x}_i^T\mathbf{n}_0 + d_0) \|\\
	   & \geq e^s_i(\alpha_0,d_0| \mathbf{x}_i)
	   -2\|\mathbf{x}_i^T\mathbf{n}\mathbf{n}-\mathbf{x}_i^T\mathbf{n}_0\mathbf{n}_0\|
	   -2\|\mathbf{n}d- \mathbf{n}_0d_0  \|
	\end{split}
	\label{eq:symmetry1-lower}
\end{equation}
\normalsize
We furthermore have
\small
\begin{equation}
\begin{split}
&\|\mathbf{x}_i^T\mathbf{n}\mathbf{n}-\mathbf{x}_i^T\mathbf{n}_0\mathbf{n}_0\| \\
& = \|\mathbf{x}_i^T\mathbf{n}\mathbf{n}-\mathbf{x}_i^T\mathbf{n}_0\mathbf{n} +\mathbf{x}_i^T\mathbf{n}_0\mathbf{n}- \mathbf{x}_i^T\mathbf{n}_0\mathbf{n}_0\| \\
&\leq |\mathbf{x}_i^T\mathbf{n}-\mathbf{x}_i^T\mathbf{n}_0|\cdot \|\mathbf{n}\|+ |\mathbf{x}_i^T\mathbf{n}_0|\cdot \|\mathbf{n}-\mathbf{n}_0\|\\
& \leq 2 \|\mathbf{x}_i\|\cdot\|\mathbf{n}-\mathbf{n}_0\| = 2\gamma_\alpha \|\mathbf{x}_i\|
\end{split}
\label{eq:sym_subs_1}
\end{equation}
\normalsize
and
\small
\begin{equation}
	\begin{split}
	&\|\mathbf{n}d- \mathbf{n}_0d_0  \|
	 = \|\mathbf{n}d-\mathbf{n}d_0+\mathbf{n}d_0- \mathbf{n}_0d_0  \| \\
	&\leq |d-d_0| +|d_0| \|\mathbf{n}-\mathbf{n}_0\|
	\leq \gamma_d + |d_0|\gamma_\alpha.
	\end{split}
	\label{eq:sym_subs_2}
\end{equation}
\normalsize
Substituting \eqref{eq:sym_subs_1} and \eqref{eq:sym_subs_2} in \eqref{eq:symmetry1-lower}, we finally obtain
\small
\begin{equation}{\label{eq:symmetry-model-lower}}
	\begin{split}
	&\underline{e}^s_i(\alpha,d| \mathbf{x}_i) \\ \doteq &\text{max}(e^s_i(\alpha_0,d_0| \mathbf{x}_i)-2(2\gamma_\alpha\|\mathbf{x}_i\|+\gamma_d + |d_0|\gamma_\alpha),0).
	\end{split}
\end{equation}
\normalsize

\textbf{Lower Bound of symmetry term} $\underline{e^s_i}(r,\mathbf{t},\alpha,d| \mathbf{y}_i)$: By using $\hat{\mathbf{n}} = \mathbf{R}^T \mathbf{n}$ and $\hat{d} = \mathbf{t}^T\mathbf{n}+d$, we analogously derive
\small
\begin{equation}
	\begin{split}
	   & e^s_i(r,\mathbf{t},\alpha,d| \mathbf{y}_i) \geq 
       e^s_i(r_0,\mathbf{t}_0,\alpha_0,d_0| \mathbf{y}_i) \\
	   & -2\|\mathbf{y}_i^T\hat{\mathbf{n}}\hat{\mathbf{n}}-\mathbf{y}_i^T\hat{\mathbf{n}}_0\hat{\mathbf{n}}_0\|
	   -2\|\hat{\mathbf{n}}\hat{d}- \hat{\mathbf{n}}_0\hat{d}_0  \|
	\end{split}
	\label{eq:symmetry2-lower}
\end{equation}
\normalsize
By substituting $\hat{\mathbf{n}} = \mathbf{R}^T \mathbf{n}$, similar to \ref{eq:sym_subs_1},  the first term gives
\small
\begin{equation}
	\begin{split}
	&\|\mathbf{y}_i^T\mathbf{R}^T\mathbf{n}\mathbf{R}^T\mathbf{n}-\mathbf{y}_i^T\mathbf{R}_0^T\mathbf{n}_0\mathbf{R}_0^T\mathbf{n}_0\|\\
	& \leq   2\|\mathbf{y}_i\| \| \mathbf{R}_0^T\mathbf{n}_0-\mathbf{R}^T\mathbf{n}\|\\
	& =  2\|\mathbf{y}_i\|\| \mathbf{R}_0^T\mathbf{n}_0 - \mathbf{R}^T\mathbf{n}_0 + \mathbf{R}^T\mathbf{n}_0-\mathbf{R}^T\mathbf{n}\| \\
	& \leq  2 \|\mathbf{y}_i\| (\gamma_\alpha+\gamma_r).
	\end{split}
	\label{eq:sym2_subs_1}
\end{equation}
\normalsize
By also substituting $\hat{d} = \mathbf{t}^T\mathbf{n}+d$, the second term gives
\small
\begin{equation}
\begin{split}
&\|  (\mathbf{t}_0^T\mathbf{n}_0+d_0)\mathbf{R}_0^T\mathbf{n}_0-(\mathbf{t}^T\mathbf{n}+d)\mathbf{R}^T\mathbf{n}  \| \\
&\leq | \mathbf{t}_0^T\mathbf{n}_0+d_0| \cdot \|\mathbf{R}^T_0\mathbf{n}_0 - \mathbf{R}^T\mathbf{n} \| \\
& + \|\mathbf{R}^T\mathbf{n}\| \|\mathbf{t}_0^T\mathbf{n}_0 -\mathbf{t}^T\mathbf{n}\| + \|\mathbf{R}^T\mathbf{n}\| |d_0 -d|\\
&\leq (\gamma_\alpha + \gamma_r) | \mathbf{t}_0^T\mathbf{n}_0+d_0|  + \|\mathbf{t}_0\|\gamma_\alpha +\gamma_t+\gamma_d,
\end{split}
\label{eq:sym2_subs_2}
\end{equation}
\normalsize
where we have used $\|\mathbf{t}_0^T\mathbf{n}_0-\mathbf{t}^T\mathbf{n}\|=\|\mathbf{t}_0^T\mathbf{n}_0-\mathbf{t}_0^T\mathbf{n}+\mathbf{t}_0^T\mathbf{n}-\mathbf{t}^T\mathbf{n}\| \leq \|\mathbf{t}_0\|\|\mathbf{n}_0-\mathbf{n}\|+\|\mathbf{n}\|\|\mathbf{t}_0-\mathbf{t}\|$. Substituting \eqref{eq:sym2_subs_1} and \eqref{eq:sym2_subs_2} in \eqref{eq:symmetry2-lower}, we finally obtain
\small
\begin{equation}
	\begin{split}
	&\underline{e}^s_i(r,\mathbf{t},\alpha,d| \mathbf{y_i})
	\doteq \text{max}(e^s_i(r_0,\mathbf{t}_0,\alpha_0,d_0| \mathbf{y}_i)-2(\gamma_t+\gamma_d \\
	& + (\gamma_\alpha +\gamma_r) \left( 2\|\mathbf{y}_i\| + |\mathbf{t}_0^T\mathbf{n}_0+d_0| \right)+ \|\mathbf{t}_0\|\gamma_\alpha),0).
	\end{split}
\label{eq:symmetry-data-lower-term}
\end{equation}
\normalsize


\section{Implementation}
Similar to prior art~\cite{yang2015go}, we improve the algorithm's ability to handle the dimensionality of the problem by installing a nested BnB paradigm.
\subsection{Nested BnB}
We install a nested BnB scheme in which the outer layer searches through the space $C_{r\alpha}$ of all angular parameters, while the inner layer optimizes over the space $C_{\mathbf{t}d}$ of translation and depth. While finding the bounds in a sub-volume of the angle space, the algorithm calls the inner BnB algorithm to identify the optimal translation and scale. One important approximation that accelerates the execution is that---whenever the bounds in a sub-volume are derived---the uncertainty of the non-optimized parameters of that particular layer are set to zero. Detailed descriptions are given in Algorithm \ref{al:algorithm1} (the Main Algorithm) and Algorithm \ref{al:algorithm2} (the Inner BnB).
\small
\begin{algorithm}[t]
\caption{Main Algorithm: BnB search for optimal registration and symmetry parameters}
\label{al:algorithm1}
\begin{algorithmic} 
\REQUIRE \emph{Data} and \emph{model} point set; threshold $\tau$; initial intervals $\mathcal{C}_{r\alpha}$ and $\mathcal{C}_{\mathbf{t}d}$.
\ENSURE Globally minimal error $E^*$ and the optimal $r^*,\mathbf{t}^*,\alpha^*,d^*$
\STATE Put $\mathcal{C}_{r\alpha}$ into priority queue $\mathcal{Q}_{r\alpha}$. Set $E^* = +\infty$.
\LOOP
    \STATE Read out interval with lowest  $\underline{E}_{r\alpha}$ from $\mathcal{Q}_{r\alpha}$.
    \STATE Quit the loop if $E^*-\underline{E}_{r\alpha} < \tau$.
    \STATE Divide interval into 4 sub-intervals.
    \FOR {each sub-interval $C_{r\alpha}$}
          \STATE Compute the corresponding optimal $\mathbf{t}$ and $d$ by calling Algorithm \ref{al:algorithm2} with $\mathbf{R}_{r_0}$ and $\mathbf{n}_0$ (zero rotation and normal uncertainty).
          \STATE Compute $\overline{E}_{r\alpha}$ and $\underline{E}_{r\alpha}$ for $C_{r\alpha} $  with the optimal $\mathbf{t},d$.
          \IF{$\overline{E}_{r\alpha} <E^*$} 
            \STATE Run ICP with initialization of $(r_0, \mathbf{t}_0, \alpha_0,d_0)$
            \STATE Update $E^*, r^*, \alpha^*, t^*$, and $d^*$ with the results of ICP.
          \ENDIF
          \STATE Discard $C_{r\alpha}$ if $\underline{E}_{r\alpha}\geq E^*$; otherwise put it into $\mathcal{Q}_{r\alpha}$
    \ENDFOR
\ENDLOOP
\end{algorithmic}
\end{algorithm}
\normalsize
\small
\begin{algorithm}[t]
\caption{BnB search for optimal translation and depth given rotation and normal}
\label{al:algorithm2}
\begin{algorithmic} 
\REQUIRE \emph{Data} and \emph{model} point set; threshold $\tau$; initial intervals  $\mathcal{C}_{\mathbf{t}d}$; Currently lowest error $E^*$
\ENSURE  Minimal error $E^*$ and the optimal $\mathbf{t}^*,d^*$
\STATE Put $\mathcal{C}_{\mathbf{t}d}$ into priority queue $\mathcal{Q}_{\mathbf{t}d}$.
\LOOP
    \STATE Read out interval with lowest  $\underline{E}_{\mathbf{t}d}$ from $\mathcal{Q}_{\mathbf{t}d}$.
    \STATE Quit the loop if $E^*-\underline{E}_{\mathbf{t}d} < \tau$.
    \STATE Divide interval into 8(2D)/16(3D) sub-intervals.
    \FOR {each sub-interval $C_{\mathbf{t}d}$}
          \STATE Compute $\overline{E}_{\mathbf{t}d}$ and $\underline{E}_{\mathbf{t}d}$ for $C_{\mathbf{t}d} $  with the  $r_0,\mathbf{n}_0$.
          \IF{$\overline{E}_{\mathbf{t}d} <E^*$} 
            \STATE Update $E^*$ and $\mathbf{t}^*,d^*$.
          \ENDIF
          \STATE Discard $C_{\mathbf{t}d}$ if $\underline{E}_{\mathbf{t}d}\geq E^*$; otherwise put it into $\mathcal{Q}_{\mathbf{t}d}$
    \ENDFOR
\ENDLOOP
\end{algorithmic}
\end{algorithm}
\normalsize
\subsection{Integration with Local ICP}
Within the outer layer, whenever BnB identifies an interval $C_{r\alpha}$ with an improved upper bound, we will execute a conventional local ICP algorithm starting from the center of the $C_{r\alpha}$ and taking $\mathbf{t}^*$ and $d^*$ as an initial value. Once ICP converges to a local minimum with a lower function value, the new value is used to further reduce the upper bound. The technique is inspired by Yang et al.~\cite{yang2013go}. 

\subsection{Trimming for Outlier Handling}
A general problem with partially overlapping point sets is that---even at the global optimum---some points may simply not have a correspondence, and should hence be treated as outliers. Although the addition of symmetry and point reflections already greatly alleviates this problem, we still add the strategy  proposed in Trimmed ICP \cite{chetverikov2005robust} for robust point-set registration. More specifically, in each iteration, only a subset of the matched data points with smallest point-to-point distances are used for motion computation. In this work, we choose a $70\%$-subset for both symmetry and registration residuals.

\section{Experiments}
We now report our experimental results on both synthetic and real data. In all our experiments, we pre-normalized the pointsets such that all points are within the domain of $[-1,1]$.  The parameter $\varepsilon$ is set to $0.5$ (cf. Section \ref{sec:domains}). 
We run experiments on both 2D and 3D data, and compare our results against the open-source implementation of Go-ICP complemented by the ransac-based symmetry detection method presented in \cite{cohen2012discovering} and applied to a fusion of the aligned point sets.
For Go-ICP, the stopping criterion $\tau$ is set to $0.001\cdot 0.7\cdot N$, and for our method it is set to $0.001\cdot 0.7\cdot(M+2N)$ where $0.7$ is the trimming ratio, $M,N$ are the number of points in $\mathcal{X}$ and $\mathcal{Y}$, respectively. For~\cite{cohen2012discovering}, the number of iterations is limited to $5000$.

\subsection{Performance on 2D Synthetic Data}
Each experiment is generated by taking an image that contains a symmetrical object, and using the Sobel edge detector to extract the object contour points. To evaluate the performance, we randomly divide the contour points into two subsets with a defined and structured overlap. To conclude, $\mathcal{Y}$ is transformed by a random rotation and translation drawn from the intervals $\pm180$ degrees and $\pm0.5$.

\begin{figure}
    \centering
	\includegraphics[width=\columnwidth]{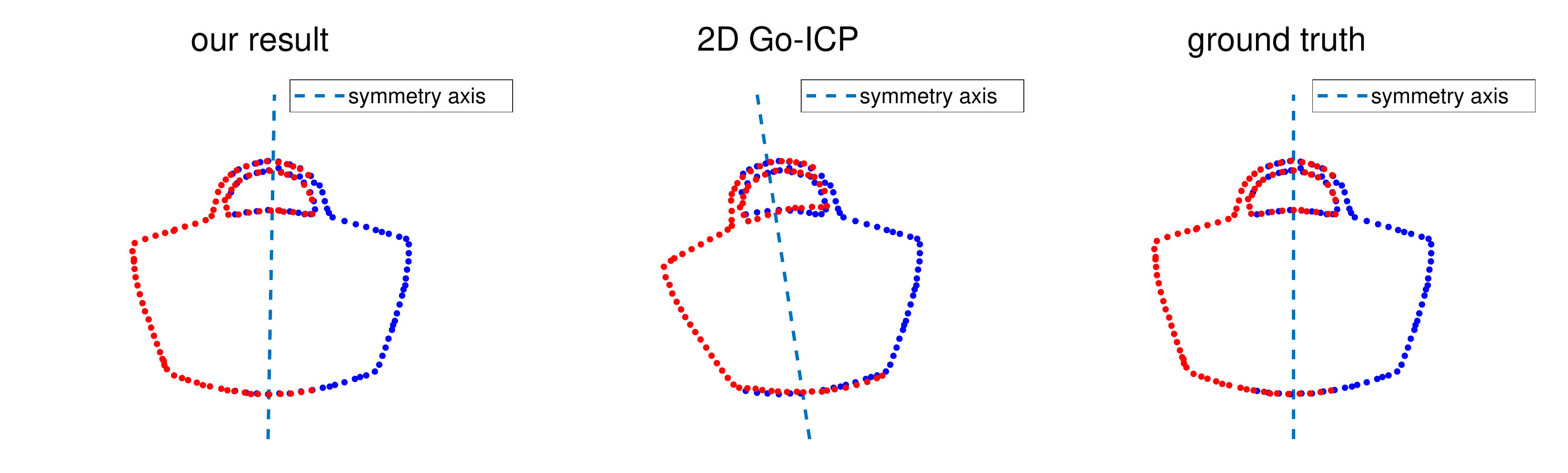}
	\caption{Example of a 2D point set registration with an overlap ratio of $0.3302$.}
	\label{fig:2d_over_visualization}
\end{figure}

\begin{figure}
    \centering
	\includegraphics[width=\columnwidth]{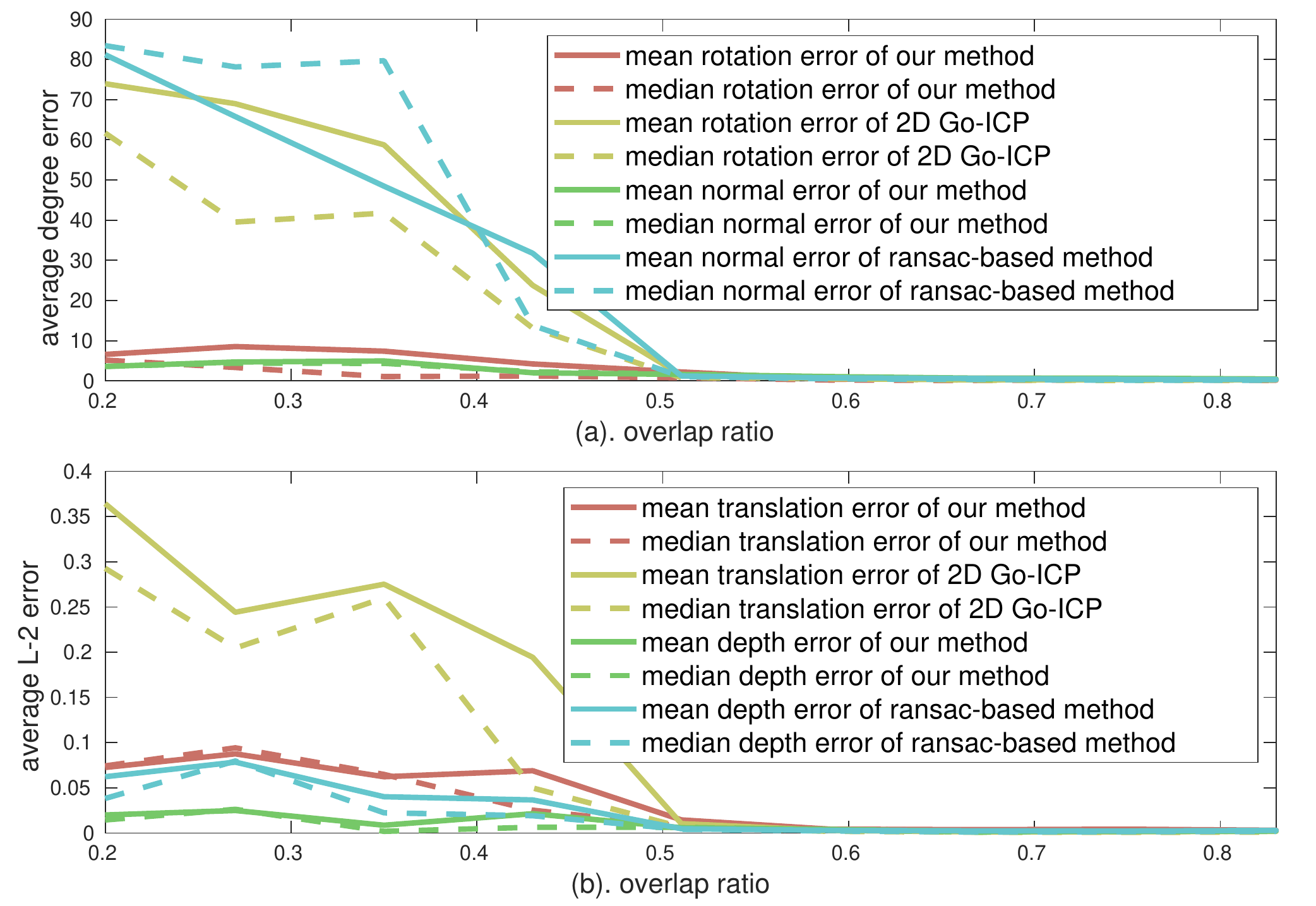}
	\caption{Mean and median errors of 2D registration compared against 2D Go-ICP followed by ransac-based symmetry detection~\cite{cohen2012discovering}. (a) shows angular errors for the rotation and the symmetry plane normal, while (b) shows the errors of the translation and the depth of plane.}
	\label{fig:2d_results_ratio_overlapping}
\end{figure}

\noindent\textbf{Handling of limited overlap}:
The overlap between both point sets if varied from $20\%$ to $83\%$. For each overlap ratio, we repeat 50 experiments each time choosing a random object, rotation and translation. Fig \ref{fig:2d_over_visualization} shows an example result, where the left  is the result of our proposed method, the centre  is the result of the 2D version of Go-ICP, and the right one shows the ground truth alignment. Figure \ref{fig:2d_results_ratio_overlapping} shows all mean and median errors of all optimisation parameters over all evaluation results. As can be observed, our method has a substantially better ability to deal with partial overlaps compared to Go-ICP, thus leading to vastly improved symmetry plane parameters as well. 

\begin{figure}
    \centering
	\includegraphics[width=\columnwidth]{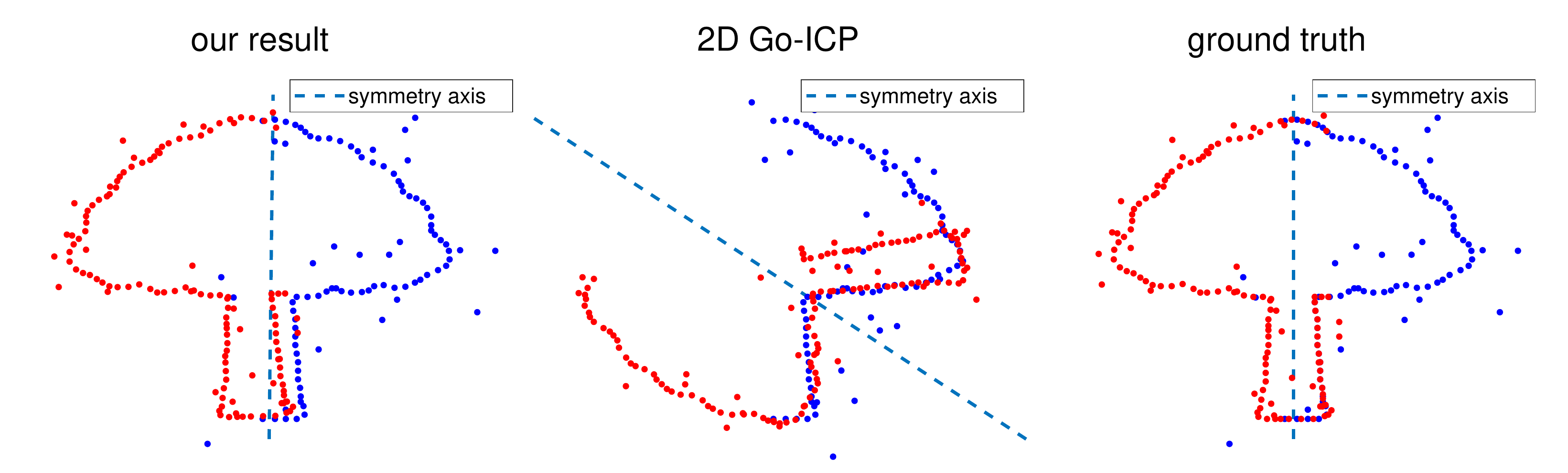}
	\caption{Example of 2D point set registration with an overlap ratio of $0.3288$ and $30\%$ outliers.}
	\label{fig:2d_outliers_visualization}
\end{figure}

\begin{figure}
    \centering
	\includegraphics[width=\columnwidth]{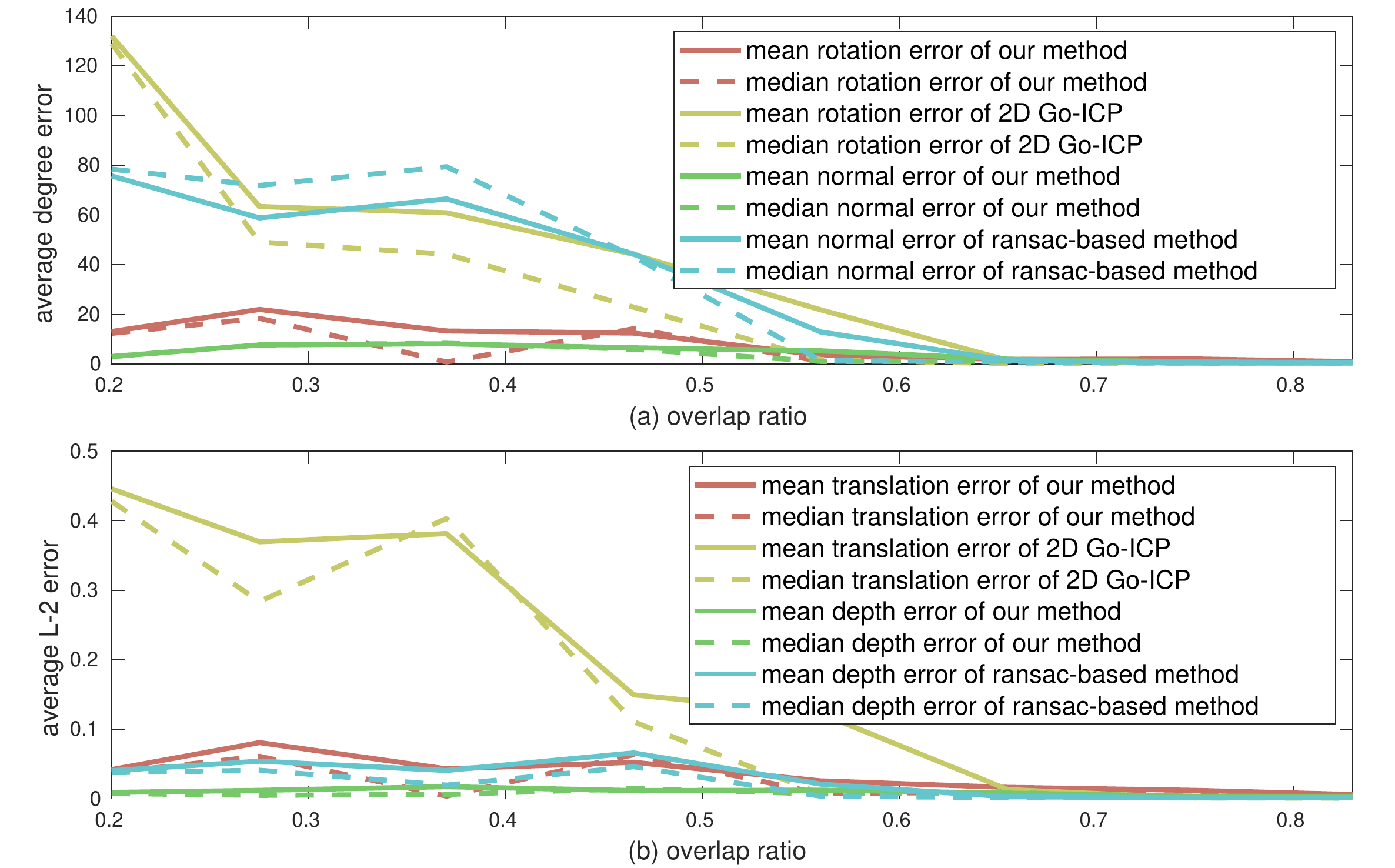}
	\caption{Mean and median errors of 2D registration using the same methods and up to $30\%$ outliers. (a) shows angular errors for the rotation and the symmetry plane normal, while (b) shows the errors of the translation and the depth of plane.}
	\label{fig:2d_results_ratio_overlapping_outlier}
\end{figure}

\noindent\textbf{Outlier handling}:
To test resilience against outliers, we repeat the same experiment but add up to $30\%$ outliers to both $\mathcal{X}$ and $\mathcal{Y}$. Figure \ref{fig:2d_outliers_visualization} indicates an example result, and Figure \ref{fig:2d_results_ratio_overlapping_outlier} again illustrates the mean and median errors over all experiments. While the registration error starts to increase earlier and the average angular errors tend to be higher, it can still be concluded that our method significantly outperforms Go-ICP followed by symmetry plane detection using~\cite{cohen2012discovering}.

\subsection{Performance on 3D Synthetic Data}

\begin{figure}
     \centering
	\includegraphics[width=0.85\columnwidth]{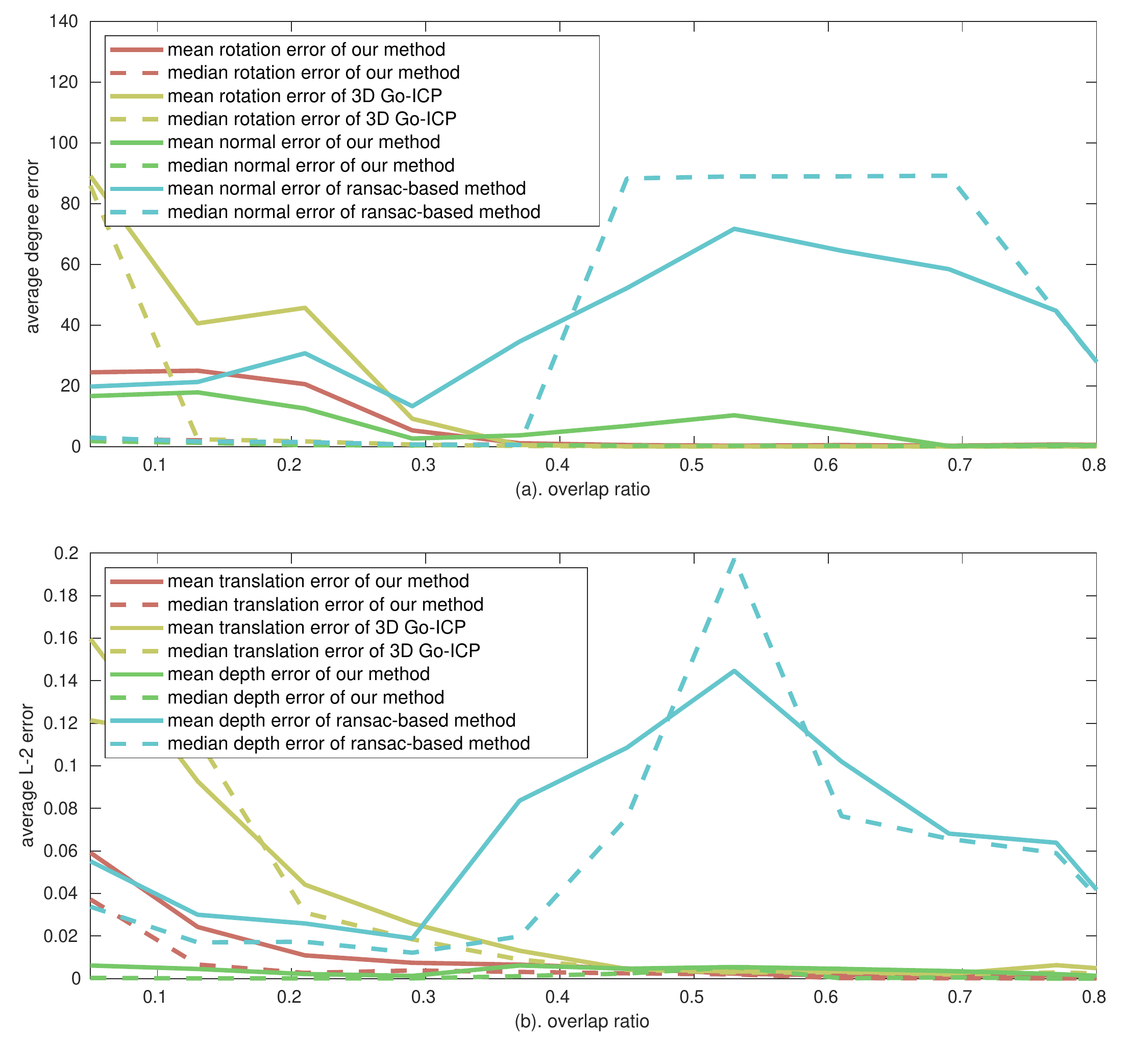}
	\caption{Mean and median errors of 3D registration compared against 3D Go-ICP followed by ransac-based symmetry detection~\cite{cohen2012discovering}. Note that no outliers are added, but--as for the case of the results in 2D--a similar trend has been confirmed for up to 30\% outliers.}
	\label{fig:3d_results_ratio_overlapping_bla}
\end{figure}

\begin{figure}
    \centering
	\includegraphics[width=\columnwidth]{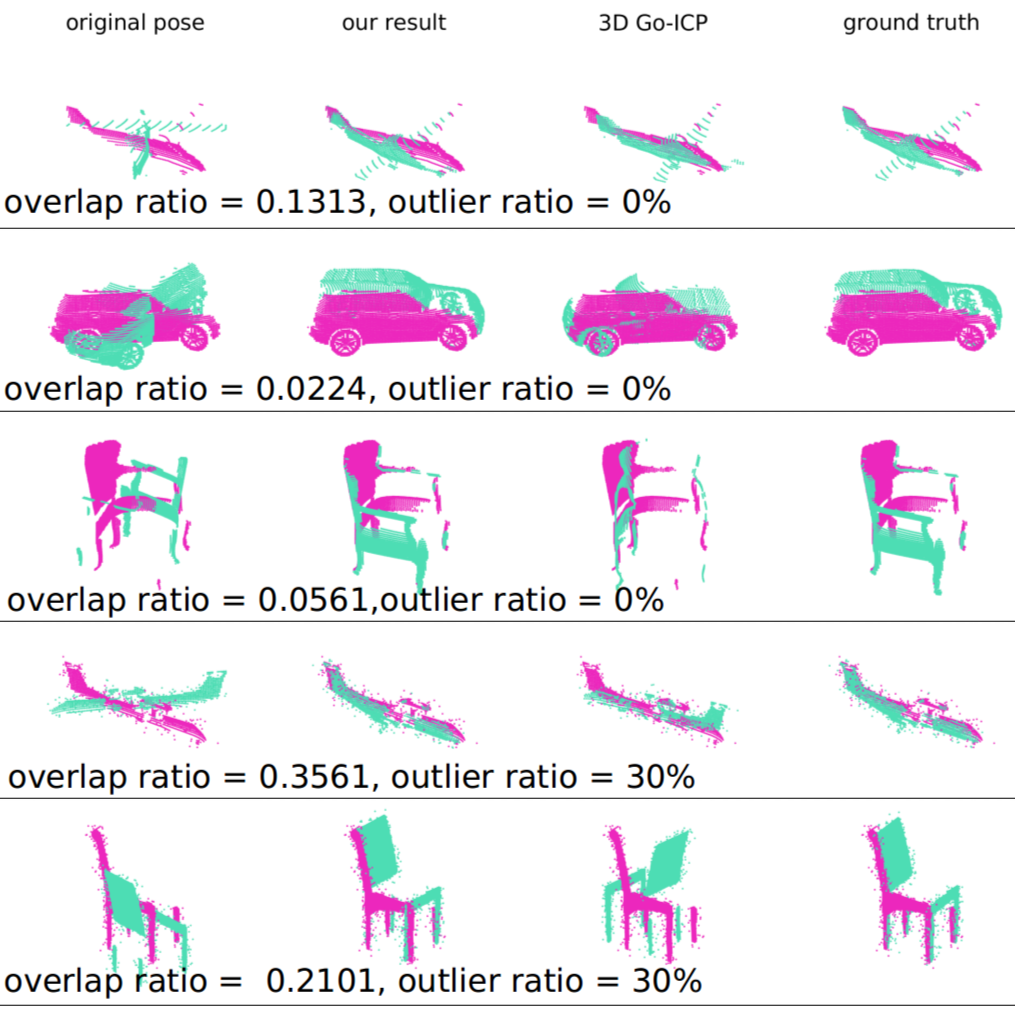}
	\caption{3D alignment results for concrete experiments. Overlap ratio and added outliers are each pair indicated.}
	\label{fig:3d_results_ratio_overlapping}
\end{figure}

\begin{figure}
    \centering
	\includegraphics[width=\columnwidth]{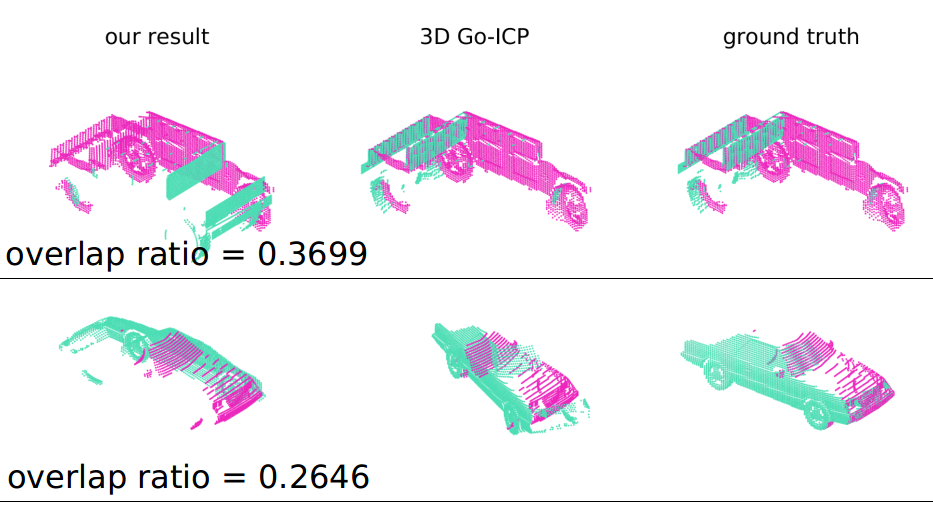}
	\caption{Example of failure cases where there is ambiguity in the symmetry plane.}
	\label{fig:failure_case}
\end{figure}

%
We choose $24$ symmetrical CAD models from ShapeNet~\cite{chang2015shapenet},  $3$ different  types from $8$ classes.  There $8$ objects contain more than one symmetry plane. For each object, we generate $16$ depth images (with occlusions) from random views around the object, and producing $109$ pairs of point sets for each object instance with varying overlap, and finally a total of about $2600$ point-set registration experiments. Our result is shown in Figure \ref{fig:3d_results_ratio_overlapping_bla}, which again illustrates mean and median errors of all estimated quantities as a function of the overlap between the sets.
In 3D, it is natural that the camera captures only a small part of the object, and that in turn even the fusion of both aligned point scans may not enable a stable estimation of the symmetry plane,
hence the slightly increased mean error.
However, especially the median error is still lower compared to Go-ICP, thus confirming the mutual benefits of our joint estimation paradigm.
Figure \ref{fig:3d_results_ratio_overlapping} shows a few concrete examples for which the 3D objects only contain a single symmetry plane and for which our method is largely outperformed. 
Figure \ref{fig:failure_case} shows failure examples where the partial point sets lead to an ambiguity in the symmetry planes. In particular, in the first example, Go-ICP still works as the overlap ratio is sufficient for the registration.
\subsection{Experiments on Real Data}
\begin{figure}
\centering
\centering
\includegraphics[width=\columnwidth]{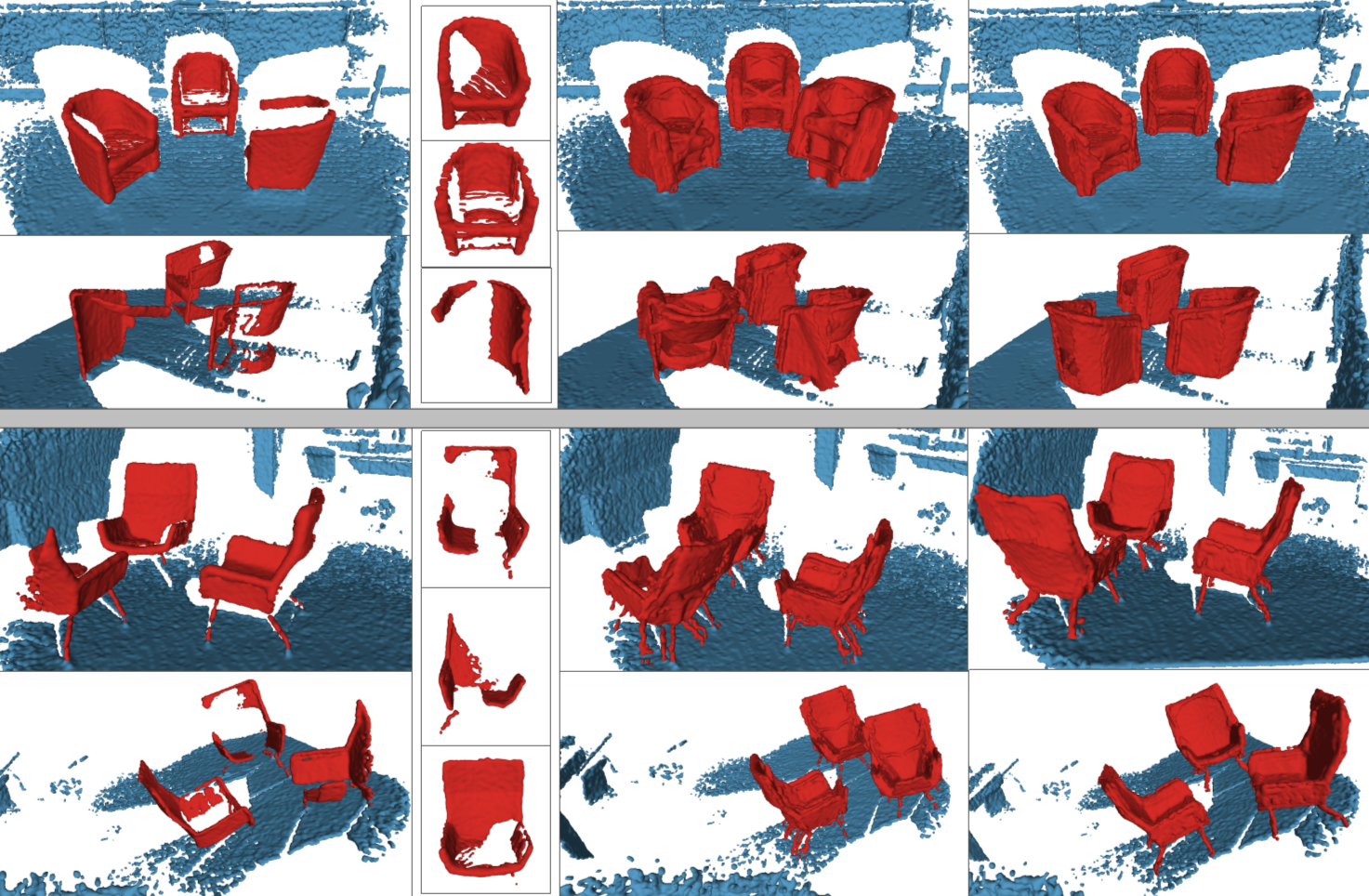}
\caption{Shape completion by alignment and transfer of segmented, partial object scans in RGBD images. The center column shows the completion results obtained by applying Go-ICP, while the right column shows the result obtained by our method.}
\label{fig:3d_real_scene}
\end{figure}

Our last experiment is an exciting application to real data that goes back to the initial motivation in the introduction. Figure \ref{fig:3d_real_scene} shows two different depth images captured by a Kinect camera, each one containing three instances of the same object under different orientations. By pairwise alignment of partial object scans, the mutual information is transferred thus leading to more complete perception of each individual object. Note that we use simple ground plane fitting and depth discontinuity-aware point clustering within object bounding boxes to isolate the partial object scans. With known position of the ground plane, we then transform the whole scene to be orthogonal to the ground plane and meet the assumption that all objects are placed upright and---in terms of relative rotation---differ only by an angle about the vertical axis. In Figure \ref{fig:3d_real_scene}, the first column shows the original scan in different orientations, the second one the partial object measurements, the third one the completion obtained by using Go-ICP as an alignment algorithm and ransac-based symmetry detection, and the last one the result obtained by using our algorithm. As can be observed, our joint alignment strategy outperforms the comparison method, and achieves meaningful shape completion.

\section{Discussion}

Symmetry detection and point set alignment over sets with small overlap are challenging problems if handled separately. Our work demonstrates a substantial improvement in both accuracy and success rate of the alignment by solving those two problems jointly. The information gained from estimating symmetry and reflecting points notably makes up for otherwise missing correspondences. However, our current implementation is not competitive in terms of running time, hence we are working on a parallel implementation. We furthermore plan to extend the algorithm to multi-point set registration, and improve its ability to deal with the situation of multiple symmetry planes.


{\small
\bibliographystyle{ieee_fullname}
\bibliography{main}
}

\end{document}